# Structural (dis)order and dynamic propensity in a mildly undercooled glass-forming liquid: Spatial correlations and the role of crystalline environments

M Shajahan G Razul[1]*, Gurpreet S Matharoo[2,3]† and Balakrishnan Viswanathan[1]‡

[1] Department of Chemistry, St. Francis Xavier University, Antigonish, NS, B2G 2W5, Canada.
[2] ACENET, St. Francis Xavier University, Antigonish, NS, B2G 2W5, Canada.
[3] Department of Physics, St. Francis Xavier University, Antigonish, NS, B2G 2W5, Canada.

*sgrazul@stfx.ca (Corresponding author)

† gmatharo@stfx.ca

‡ bviswana@stfx.ca

## ABSTRACT

We use the isoconfigurational (IC) ensemble to show the connection between emerging heterogeneities in the tetrahedral order parameter and the dynamic propensity in a mildly undercooled glass-forming liquid. We observe that spatially correlated tetrahedrally-(dis)ordered clusters of molecules are observable on the time scale of structural relaxation. The heterogeneities of tetrahedrally-(dis)ordered clusters correlate with dynamical heterogeneities (DH) and these correlations reach peaks at similar time scales. We discover that the angular component of the tetrahedral order parameter is strongly correlated to the dynamics compared to the radial component. Moreover, these correlations between the dynamics and tetrahedrally-(dis)ordered regions enormously influence the system, with spatial correlations being observable for a prolonged period beyond the peaks of maximum DH. Further, we discover that the crystalline particle environments in our water model (as identified by the IC ensemble) may be the origin for slow dynamics of dynamical heterogeneity in our undercooled model water system.

# 1. INTRODUCTION

Liquid water displays many interesting properties at ambient temperature[1] and further displays many anomalies[2,3] when it remains liquid below its freezing temperature. At supercooled temperatures (in the absence of quenching), nucleation events induce crystal formation[4] in water and — when quenched appropriately — the resultant slow dynamics usually leads to glass formation[5]. In supercooled water, the interplay between structure and dynamics is responsible for the many anomalies that occur[5] with a decrease in temperature and is the focus of much research work[2,5,6]. It is known that dynamical heterogeneity arises close to the glass transition[5]. Dynamical heterogeneity refers to the mobile and immobile particles in the system showing preference to be in spatial regions that are distinctly populated by the slowest particles and other regions by the fastest particles, in contrast to these particles scattered randomly throughout the system. Specifically, in glass forming liquids, α relaxation, $\tau_\alpha$ increases upon supercooling. Further, dynamical studies[7] have shown that the slowing down of dynamics is consistent with spatial regions[8] with dynamics that vary drastically, which is consistent with the idea of heterogeneity of glassy relaxation[9]. Adam and Gibbs[10] have postulated that cooperatively rearranging regions exist. The molecules in these regions possess configurations that change independently of the whole system (and other similar regions); these regions have been observed to grow in size with a decrease in temperature close to the glass transition temperature, exhibiting a characteristic length scale[11,12]. A definitive explanation of how these length scales play a determining role in restructuring and concurrent structural relaxation is still actively investigated[13,14]. Of note are studies[15,16] that have shown that changes in the tetrahedral network of water occurs as the temperature is decreased is linked to the dynamics of metastable forms of water, establishing a link between the dynamics and structure of these states of liquid water close to the glass transition[16]. However, there is still a lack of consensus on the connections between the dynamic properties of water, its associated dynamical heterogeneity and changes in the local structural properties[17]. Authors such as Tanaka[18] have also strongly suggested a link between the glass transition and crystallization, since slow dynamics in a supercooled liquid can be associated with both processes.

One important tool to aid in elucidating these questions is the isoconfigurational (IC) ensemble[19], initially proposed by Harrowell and co-workers. The IC ensemble is composed of $N$ independent microcanonical simulation runs, all beginning from an identical configuration; however, the momenta are scrambled randomly and assigned values from the Maxwell-Boltzmann distribution appropriate for the temperature required. The distribution of the system-averaged mean squared displacement for each particle (across the $N$ runs) is the particle realization of propensity for movement from its initial configuration. The IC ensemble has also been used to investigate coarse-grained polymers[20] and a glass-forming polymer[21]. These studies and other previous work on a water model[22] and a model glass former[23] have definitely shown that the IC ensemble offers a crucial and unique insight: it exposes in real space the morphology of the structural heterogeneity present in a single instantaneous configuration. This information is typically lost, when the conventional average over initial configurations is employed. In these studies, it has been further demonstrated that the correlations exposed *via* IC averaging are properties of the initial configuration (from which the runs of the IC ensemble are generated) that have been obtained by an equilibrated system. It is the sensitivity of the dynamics to the initial configuration that reveals the spatial heterogeneity observed in the (dynamic) propensity and at least in the case of the model glass former[23] and glass forming polymer[21] the local structure and dynamical propensity are correlated. All these studies investigated the peaks of various correlations to identify the dynamical heterogeneities and it is still unclear if the dynamically revealed static correlations as revealed by the IC averaging persisted spatially after the various correlation measures became uncorrelated in these glass-forming liquids.

Here, we investigate the connection between structure (particularly the tetrahedral structure) and dynamics of water just below the freezing temperature using a model water system with SPC/E[24] at 265 K. The diffusion coefficient of SPC/E at this temperature is the same as that of supercooled water reported experimentally at -2 °C[25]. SPC/E at 265 K was selected to match a mildly undercooled cooled system to determine how long the dynamical heterogeneity persists in the mildly undercooled liquid from an arbitrary initial configuration as it has not been adequately addressed in previous studies. Moreover, it allows characterization of the emergence and disappearance of these dynamically heterogeneous regions, unlike more deeply supercooled water which requires much longer simulation times. Furthermore, the use of tetrahedral order will be used to identify crystalline environments[26] and we can demonstrate the connections between crystal-like particles and dynamical heterogeneity in water. We aim to show that heterogeneities of the tetrahedral order parameter exist, where the spatial regions of high and low tetrahedral order exist and are linked to the underlying dynamical heterogeneity, as opposed to being random and scattered evenly (spatially) throughout the system. We thus employ the IC ensemble to uncover the properties of this water system to highlight its utility as a tool and for the important structural and dynamical insights it can provide beyond single configurations.

## 2. METHODOLOGY

### 2.1 Model and simulation details

We carried out classical molecular dynamics (MD) simulations of 1000 molecules of SPC/E water at $T = 265$ K and $\rho = 0.99$ g/cm$^3$ under *NVE* conditions with 500 IC simulations for 1 ns to generate the data presented here. All MD simulations were carried out utilizing the GROMACS package[27]. The time step was 2 fs in an isotropic cubic simulation box with periodic boundary conditions. The system was created utilizing the tools available in GROMACS and equilibrated at 300 K over 20 ns under *NPT* conditions (pressure: 1 atm) using a Berendsen thermostat and barostat[28]. The system was then further equilibrated at 265 K under *NPT* conditions for 50 ns using a Parinello-Rahman barostat[29] and a Nosé-Hoover thermostat[30]. Equilibration was carried out in 10 ns intervals with the time duration of thermostat - barostat coupling increased during this interval. This equilibrated system was used as the starting point for the IC ensemble runs under the microcanonical ensemble. A cut-off radius of 0.9 nm was used to treat non-bonding interactions and the particle mesh Ewald method[31] was used for long-range electrostatic interactions. Two independent simulations and two random equilibrated configurations were used to confirm the results presented. The results of two independently equilibrated averages of the IC configurations are presented in this work.

Determination of tetrahedral order for individual configurations and IC ensemble analysis were performed using the tetrahedral order parameters of Chau and Hardwick[32]. These provide separate but mutually important data of the bond angles and bond distances of the four nearest neighbors. These have been used to definitely identify crystalline particles in steady-state crystallization studies of water[26] at melting and freezing ice-water interfaces. The first expression is concerned with the angular dependence of the arrangement of the four nearest neighbors,

$$Sg = \frac{3}{32}\sum_{j=1}^{3}\sum_{k=j+1}^{4}\left(\cos\psi_{j,k} + \frac{1}{3}\right)^2 \qquad (1)$$

where $\psi_{j,k}$ is the angle subtended at the central atom between the $j^{th}$ and $k^{th}$ intermolecular separation vectors (bonds), the factor 3/32 normalizes $Sg$ to the range $0 \leq Sg \leq 1$, and squaring ensures that the contribution from each bond angle is always $\geq 0$. The tetrahedral arrangement of all the angles between bonds is the same, and the cosine of the angle is - 1 /3. The double sum ensures that all six angles are accounted for in calculating the Sg function. The second part focuses on the distances of the four nearest neighbors and is given by

$$Sk = \frac{1}{3}\sum_{k=1}^{4}\frac{(r_k - \bar{r})^2}{4\bar{r}^2} \qquad (2)$$

where $r_k$ is the radial distance from the central molecule to the $k^{th}$ neighbor, $\bar{r}$ is the arithmetic mean of the four radial distances, and 1/3 the normalizing factor. Both $Sg$ and $Sk$ return zero for a perfectly tetrahedral arrangement of the four nearest neighbors and a maximum value of 1 for complete deviations from the angular and distance ideals. Chau and Hardwick have also provided $Sg$ and $Sk$ distributions of water and ice and the average values are also provided. These order parameters define tetrahedral environments or local crystalline environments for a water molecule and its four nearest neighbours. Utilizing the IC ensemble we can define the dynamic propensity[22] of each oxygen atom in water as

$$\langle r_i^2 \rangle_{ic} = M^{-1} \sum_{k=1}^{M} r^2(i,k,t) \qquad (3)$$

where $i$ is the particle, $k$ the run of the IC ensemble, $t$ the time when the averaging is performed, and $M$ the number of runs in the ensemble. In this work we also obtain the IC-averaged tetrahedral order parameter; the *Sg* and *Sk* values are defined as

$$\langle St_i \rangle_{ic} = M^{-1} \sum_{k=1}^{M} St(i,k,t) \qquad (4)$$

where $St = Sg$ or $Sk$ and all other parameters are as described previously. Here, the parameter indicates the likelihood of forming tetrahedrally ordered or disordered environments and aids in the identification of crystalline environments around molecules.

### 2.2 Cluster analysis

Cluster analysis[22] is used to investigate spatial correlations of the lowest and highest 5%[33] of values of $\langle r_i^2 \rangle_{ic}$, $\langle Sg_i \rangle_{ic}$ and $\langle Sk_i \rangle_{ic}$ of oxygen atoms only. This is accomplished by defining a cluster of two oxygen atoms if they are within a distance of 0.32 nm (the first minimum of the O-O radial distribution function, *g(r)*). The initial configuration serves as the coordinate space upon which cluster analysis is performed. We then define the number-averaged mean cluster size of a set of $N_c$ clusters, within a time interval, $\Delta t$ as

$$S_n(\Delta t) = \frac{1}{N_c} \sum_n nN(n, \Delta t). \qquad (5)$$

where *N(n, Δt)* is the number of clusters of size *n*. $S_n(\Delta t)$ then measures the number-average size of a cluster over a distribution of clusters of varying sizes over the time interval. In order to eliminate the random contribution of a particle to the cluster for the property we are measuring we define the normalized number-average size of a cluster,

$$\overline{S_n}(\Delta t) = \frac{S_n(\Delta t)}{S_r(\Delta t)} \quad (6)$$

where, $S_r(\Delta t)$ is the value that results when 5% of the particles are selected at random, over the same time interval regardless of the property being calculated. We find small differences to the cluster sizes if we vary the fraction of the particles chosen from 5% - 10%. We thus evaluate $\overline{S_n}(\Delta t)$ for clusters of immobile and mobile particles of $\langle r_i^2 \rangle_{ic}$. We conduct the same analysis on $\langle Sg_i \rangle_{ic}$ and $\langle Sk_i \rangle_{ic}$ and find the mean normalized number-average size of a cluster.

**2.3 Investigating Correlations**

The IC correlations that exist between high and low values of $\langle r_i^2 \rangle_{ic}$, $\langle Sg_i \rangle_{ic}$ and $\langle Sk_i \rangle_{ic}$ can be determined by Pearson's correlation coefficient, *R*. This method can be simply used to discover the strength of a link between two sets of data:

$$R(X, Y) = \frac{cov(X, Y)}{\sigma_X \sigma_y} \quad (7)$$

or written as,

$$R(X,Y) = \frac{\sum_{i=1}^{n}(X_i - \bar{X})(Y_i - \bar{Y})}{(n-1)\sqrt{\frac{\sum_{i=1}^{n} x_i^2 - n\bar{x}}{n-1}}\sqrt{\frac{\sum_{i=1}^{n} y_i^2 - n\bar{y}}{n-1}}} \quad (8)$$

$X$ and $Y$ are distinct values made of $n$ points for $x_i$ and $y_i$, with $\bar{x}$ and $\bar{y}$ the respective means respectively. The use of the correlations presented here shows that value for the strongest correlation is 1, 0 for no correlation and -1 for anti-correlation. The particle IC ensemble properties can be easily investigated for correlations for low and high values between $\langle r_i^2 \rangle_{ic}$, $\langle Sg_i \rangle_{ic}$ and $\langle Sk_i \rangle_{ic}$ in time. Therefore, strong correlations are revealed in this manner for propensity-averaged properties compared to single configuration correlations. For example in Fig. 1 a comparison is made between the correlation between IC averaged immobile particles and the IC averaged ordered $Sg$ tetrahedral order parameter. This is contrasted against the non-IC (or single configuration) of the same properties (immobile and ordered $Sg$). It is clearly demonstrated the IC averaged properties show larger and distinct correlations. This is further discussed in section 3.2

## 3. RESULTS & DISCUSSION

### 3.1 Mean Squared Displacement

We note from Fig. 2 that the mean squared displacement (MSD) shows little interesting behavior except for a hint of caging behavior before the MSD is in the diffusive regime, after only about 2 ps. The non-Gaussian parameter[34] (NGP) has been used to quantify heterogeneities in the dynamics in supercooled liquids[9,13,22]. From the results of Sciortino et al.[9] we can obtain the peak of the NGP for systems simulated at 258.5 K and 284 K. We can extrapolate the very low value of the NGP peak to be at approximately 2 ps and the NGP vanishes at 20 ps, indicating that heterogeneities exist, albeit short-lived. To further investigate the weak dynamical heterogeneity indicated by the NGP, we look at the correlations between the structure factor and the mobility of the molecules using IC ensemble averaging in the discussion below.

**3.2 Tetrahedral bonding and heterogeneous dynamics**

For better statistics of the averages obtained in Fig. 3, we averaged two independent IC starting configurations. Fig. 3a (curve *a*) shows that the correlation between immobile molecules, (*IM*) $\langle r_i^2 \rangle_{ic}$ and tetrahedrally (angular) ordered, (*O*) $\langle Sg_i \rangle_{ic}$ reaches a peak at 4 ps. The correlation of mobile molecules (*M*) $\langle r_i^2 \rangle_{ic}$ and tetrahedrally (angular) disordered, (*D*) $\langle Sg_i \rangle_{ic}$ also shows correlations, but lower in value, as shown in Fig. 3a (curve *b*). The correlations between *IM* $\langle r_i^2 \rangle_{ic}$ − *O* $\langle Sk_i \rangle_{ic}$ (curve *c*) and between *M* $\langle r_i^2 \rangle_{ic}$ − *D* $\langle Sk_i \rangle_{ic}$ (Fig. 3a curve *d*) are also lower than *IM* $\langle r_i^2 \rangle_{ic}$ − *O* $\langle Sg_i \rangle_{ic}$. The results in Fig. 3a shows that the dynamically slow particles with the highest angular tetrahedral order are more significant than the other correlations, yet these are generally much larger than has been appreciated and noted in previous work[12,35]. These correlations are much larger that in any random single configurations for the same properties as observed in Fig. 1. We also discover that an anti-correlation exits between the immobile domains and disordered *Sg* and *Sk* values, and between mobile domains and ordered *Sg* and *Sk* values. The degree of anti-correlation occurs to a similar extent as the correlation. In Fig. 3b the $\langle Sg_i \rangle_{ic}$ − $\langle Sk_i \rangle_{ic}$ show correlations up to 10 ps, with a significant peak between 1- 2 ps. This indicates — at least from the correlations — that the structure factor is important over the same timescale as the dynamic heterogeneity in the system. The significance of this result is that there are crystalline particle environments present in the system from 2-6 ps. We observe that the non-IC $Sg_i$ and $Sk_i$ are correlated at about 30% for single configuration trajectories. The correlations highlighted in Fig. 3a and Fig. 3b appear insignificant just after 10 ps. In Fig. 3c we compare cluster sizes and find the single run peak of the *IM $r_i^2$* gives a mean cluster size, $\overline{S_n}(\Delta t)$ = 1.50 (not shown), which can be compared to the peak in the IC ensemble cluster, $\overline{S_n}(\Delta t)$ = 5.98. However, on the timescale of structural relaxation a maximum occurs, indicating clustering of immobile particles (*IM* $\langle r_i^2 \rangle_{ic}$) and to a lesser degree mobile particles (*M* $\langle r_i^2 \rangle_{ic}$). At large *t*, heterogeneities begin to dissipate, and $\overline{S_n}(\Delta t)$ decreases to 1. Unsurprisingly, clusters of disordered $\langle Sg_i \rangle_{ic}$ and $\langle Sk_i \rangle_{ic}$ are not significant and within the random limits (not shown). The mean clusters of ordered $\langle Sk_i \rangle_{ic}$ are also less significant (not shown). However, $\langle Sg_i \rangle_{ic}$ clusters are larger with a peak value of $\overline{S_n}(\Delta t)$ = 1.32 at 4 ps. The configurations revealed no other significant clusters of ordered and disordered *Sg* and *Sk* values. In crystalline particle environments both $\langle Sg_i \rangle_{ic}$ and $\langle Sk_i \rangle_{ic}$ particles should have the lowest values. It is clear from the data that no ice crystalline clusters have formed, since there are no correlations of cluster formation of these structural order parameters. Yet, we observe clustering of high tetrahedral environments, as captured by the *Sg*. It is interesting that spatial correlations exist (Fig. 3a and Fig. 3b) and the mean cluster sizes of *IM* $\langle r_i^2 \rangle_{ic}$ are especially prominent between 2-20 ps. In contrast, *IM* $\langle r_i^2 \rangle_{ic}$ clusters are very small and the *O* $\langle Sg_i \rangle_{ic}$ persists up to 10 ps. Figures 3a-c do not provide any indication where these correlations occur in the system or if they occur in the same spatial regions. We use data visualization to investigate how these spatial regions are correlated spatially in the next section.

**3.3 Data Visualization**

For the visualization of the clustering of spatial regions, Matharoo *et al.*[22] and Razul *et al.*[23] employed the method shown in Fig. 4. We carried out the method as described in Fig. 4 (please see Figure for a description of how the 3D Spatial plots are generated) for a single configuration and the same configuration via IC averaging. The single configurations show that the heterogeneity cannot be visually seen over the trajectory 2-10 ps, with the most immobile particles represented as the largest blue spheres. Similarly the most immobile atoms of $\langle r_i^2 \rangle_{ic}$ are the largest blue spheres as described in Fig. 4. In contrast to the single configuration equilibrated system, the heterogeneity is clearly observed in the IC averaged system. Fig. 4 illustrates simply the use of the visualization procedure to observe heterogeneities in mobility and the tetrahedral order parameter described in Fig. 5. In the top left panel, in Fig. 5a, we visualized the spatial variation of the dynamic propensity for both the immobile and mobile values for $\langle r_i^2 \rangle_{ic}$ for one starting configuration at $t = 4$ ps corresponding to approximately, the maximum cluster size where the clusters are prominent. Particles in the top (bottom) 50% $\langle r_i^2 \rangle_{ic}$ values are represented as blue (yellow) spheres, with each sphere plotted at the position of the particle in the initial configuration. The radius of each sphere represents the rank order of $\langle r_i^2 \rangle_{ic}$: the larger the blue (yellow) sphere is, the smaller (larger) is its value of $\langle r_i^2 \rangle_{ic}$ corresponding to immobile (mobile) regions. In Fig. 5b we visualize the spatial variation of the $\langle Sg_i \rangle_{ic}$ high and low values of tetrahedral order at t = 4 ps corresponding to the maximum cluster size where the clusters are most prominent. These are represented in a similar manner as in Fig. 5a. The radius of each sphere represents the rank order of $\langle Sg_i \rangle_{ic}$: the larger a red (green) sphere is, the smaller (larger) is its value of $\langle Sg_i \rangle_{ic}$ corresponding to tetrahedrally (angular) ordered (disordered) regions. We can observe that the regions here are distinct, perhaps not as prominent as those in References [22,23]; however, a clear connection can be observed in the same regions from inspection of the top panels, showing clear spatial correlation in regions where $IM \langle r_i^2 \rangle_{ic}$ is correlated with $O \langle Sg_i \rangle_{ic}$ and $M \langle r_i^2 \rangle_{ic}$ is correlated with $D \langle Sg_i \rangle_{ic}$. Here we observe that the correlations are in similar regions visually. In previous work, dynamical heterogeneous regions were always obtained at the peak positions of the correlations. However, Fig. 5c shows heterogeneities at 30 ps and Fig. 5d shows the same configuration at 50 ps where the heterogeneities seem to be fading; however, the regions are still visually discernible. Of significance is the persistence of these regions beyond all the other correlations in Fig. 3 and the heterogeneities described previously at 4 ps. This result is the first of its kind to show persistence of such regions beyond typical dynamical heterogeneity measures. This perhaps explains the persistence of the clusters seen in the cluster analysis (Fig. 3c) of immobile regions. It should be noted that Fig. 5a shows the regions of dynamical heterogeneity of the most mobile and immobile regions and that any clusters calculated via the cluster analysis are located in those regions; thus Fig 5a does not show clusters as defined in section 2.2. This result is also significant as it indicates that a typical equilibrated configuration under mildly undercooled conditions has regions of mobile and immobile regions that persist beyond 50 ps. Comparing Fig. 5b and 5d, it is clear that the regions that are immobile (mobile) at 50 ps occurs in the regions of tetrahedral order (disorder) at 4 ps of the $\langle Sg_i \rangle_{ic}$. The angular part of the tetrahedral order parameter indicates that ordered crystalline-like bond angles are more important due to hydrogen bonding than the radial distances of the four nearest neighbors. In steady crystallization studies[26] of water there is clear coupling of both low $Sg$ and $Sk$ values at ice-water interfaces. It is also noteworthy that separating out the angular and radial parts of the tetrahedral order parameter is important in understanding the dynamical heterogeneity of water, where true crystalline cluster environments (low $Sg$ and $Sk$ values) are not formed, only distorted crystalline cluster environments. In addition, we have observed that there are no correlated regions that coincide with the dynamic propensity and tetrahedral order at 50 ps. The IC ensemble, in its typical representation and analysis, captures propensities from the initial configurations, but not particles that have changed positions since the start of the IC ensemble trajectory. Conceivably, particles that are new to fast and slow domains may have been overlooked. Fig. 5d suggests that this may be the case, which explains the persistence of spatial heterogeneity of particles that exists longer (up to 50 ps), as seen in this work. Cluster analysis shows cluster sizes consistent with previous studies for the mobile regions[12,34]; however, the temperature is much higher in this study and the mean cluster size of immobile regions is also much larger.

The results demonstrate that the regions (or domains) present at 4 ps for both propensity and tetrahedral order influence the structure at later times (especially the regions of immobile or mobile regions). We can thus establish here that the structure and dynamics that show maximum heterogeneity at an earlier time influence dynamical heterogeneity at a later time. This strongly suggests that the IC ensemble may be a useful tool for investigating critical fluctuations of structure (from a starting configuration) that may be responsible for (for example) nucleation and subsequent crystallization events.

## 3.4 Propensity of individual atoms versus propensity on the IC ensemble trajectory

To further clarify beyond the visual representation in Fig. 5, we determined the correlation between dynamic propensities of the atoms obtained at individual time points and compared it to propensities on the IC ensemble trajectory in time using Pearson's correlation method in Fig. 6. The propensities at 0.2 ps are in the ballistic region and 0.5 ps are in the weakly caging regime, and the correlations are observed to fall off to well under 30% at 50 ps; the correlations are still present despite being weaker. However, for the propensities at 2, 4 and 6 ps, the correlations are well above 30% at 50 ps and heterogeneities are clearly seen in the 3D-spatial plots. The dynamic propensities at 0.2 ps and 0.5 ps show that the propensities in the ballistic and (weakly) caging regions of the MSD influence the subsequent dynamics overall, but weakly. However, the slight increase in both the 0.2 ps and 0.5 ps data at 20 ps show that some features are strong enough that it manifests at a later time, and which originated from the starting configuration. The same feature is readily observed at the same time point of the propensity at 2 ps and weakly at 4 ps. This feature implies that some correlations in the original configuration that are significant are just masked by the later configurations that show stronger correlations and these may vary from different starting configurations. In addition, it is clear that the propensities in the early diffusive regions affect the dynamical heterogeneity until about 100 ps (not shown), where the correlations are still just above 10% (considered weak correlations). The correlations diminish to well below 10% (considered negligible correlations) at 300 ps (not shown). The reason for the difference in correlation between the different propensities is the initial randomization of velocities in the initial configuration. In ballistic and caging (in this case weakly caging) regions (Fig. 2), the configuration already contain domains that are heterogeneous in terms of mobility, however the individual molecules locally are structurally relaxing due to the domains they are found in. For example, a molecule in an immobile domain may be randomly assigned a higher velocity; the velocity will relax to a lower velocity and *vice versa*. Therefore, these molecules will change during each time step in the IC trajectory for that molecule in the ballistic and caging regions of the MSD. This explains the dramatic de-correlation at 0.2 and 0.5 ps. However, in the initial diffusive regime — past the peak of the NGP — the configurations are more alike up to 50 ps, especially the immobile domains, which change less than the mobile domains, explaining the correlation trend seen in Fig. 6. It is likely that the immobile molecules change far less than the mobile molecules, and this is responsible for the correlations observed in Fig. 6. Regions of spatial heterogeneity (regions of mobile and immobile domains) indicate that the lowest 30% at 50 ps are still moderately correlated at 0.34 and the highest 30% (most mobile) are weakly correlated at 0.19 (not shown). The evidence here indicates that the mobile domains de-correlate faster than slow domains.

The implication for a mildly undercooled system with diffusion coefficient of $1.05 \times 10^{-5}$ cm$^2$/s is that this value coincides with water at -2 °C[25]. That such dynamical heterogeneities exist under such conditions is surprising and underappreciated. The impact of such heterogeneities at such conditions may be important in aqueous systems with the inorganic solutes and biomolecules that are important in food science[36] and cryobiology[37].

**3.4 Possible origin of the dynamical heterogeneity in water**

The IC ensemble typically exposes dynamical heterogeneity but not its origin. The propensity is observed as a response to the starting configuration, especially where dynamical heterogeneity exists in a supercooled liquid. In Fig. 7a we show the top 10% of immobile oxygen positions in the same visualization scheme presented earlier (Section 3.3) at 4 ps. For clarification we omit all other particles. The regions highlighted (circled in red) show extended regions of two regions, *A* and *B* in the simulation box. We then selected the particles with both the top 10% propensity (*IM* $\langle r_i^2 \rangle_{ic}$) and top 10% high crystalline order ($O \langle Sg_i \rangle_{ic}$) at 4 ps, shown in Fig. 7b. We observe similar particles in the same regions *A* and *B*. We next examined the top 10% with both $O \langle Sg_i \rangle_{ic}$ and $O \langle Sk_i \rangle_{ic}$ — which defines particles with high crystalline environments — and we discover that these particles first appear at 1 ps (shown in Fig. 7c) in the regions *A* and *B*. The spatial 3D plots in Figs. 5 and 7 also reveal that these are not strictly particle-for-particle correlations over time but that these particles are in the same spatial domains. These particles with high crystalline environments may be responsible for influencing the immobile regions that develop later as high angular tetrahedral order develops in these regions. What we observe here are the most ordered particles in terms of most perfectly tetrahedral (angular) particles are distributed together in the same domains before the onset of immobile dynamical heterogeneity, which is a new insight into dynamical heterogeneity. For the establishment of this as a potential mechanism for dynamical heterogeneity in this system, better statistics at this temperature with multiple initial equilibrated configurations must be undertaken before any clear conclusions may be drawn. However, this result demonstrates that the possible origin of the dynamical heterogeneity of the immobile regions in water is at least in part due to the crystalline particle environments that influence the subsequent dynamical heterogeneity at 4 ps and beyond. The data does not show such behavior for mobile particles with non-crystalline like values. The ordered *Sg* and *Sk* values indicate crystalline particle environments (high tetrahedral order) are moderately correlated to slow dynamical regions and disordered to mobile regions as revealed by the IC ensemble.

## 4. CONCLUSION

We have shown that identifiable dynamical heterogeneities do coincide with regions of high tetrahedral order as shown by the angular component — but not the radial part — of the tetrahedral order parameter in a mildly undercooled water system. The maximum correlation of dynamic propensity and tetrahedral order coincides at 4 ps and influences the clearly defined dynamical heterogeneous regions up to 50 ps. These heterogeneities persist beyond 50 ps and only appear to dissipate at 100 ps. We employed cluster analysis, a simple correlation scheme, and 3D-spatial plots to demonstrate the heterogeneities. One of the significant results communicated here is that pervasive heterogeneous domains are clearly seen — which may have been underestimated previously — beyond typical metrics for dynamical heterogeneity, as revealed by the IC analysis. It has also been shown that the possible origin of dynamical heterogeneity from the slow domains lies in high crystalline single particle domains, as revealed by both components of the tetrahedral order parameter used in this study, and we identify that this occurs 3 ps earlier. We expect the effects demonstrated in this work will be more pronounced as the temperature decreases. In addition, we characterize the IC ensemble method by showing how the propensities in the initial diffusive region influence the heterogeneity later in the simulation compared to the propensity in the ballistic regime.

## ACKNOWLEDGMENTS

This research was enabled in part by support provided by ACENET (www.ace-net.ca) and Compute Canada (www.computecanada.ca).

## CREDIT AUTHOR STATEMENTS

**M. Shajahan Gulam Razul**: Conceptualization, Formal analysis, Investigation, Methodology, Project Administration, Software, Validation, Visualization, Writing – orignal draft, review & editing **Gurpreet S Matharoo** Investigation, Validation, Writing- review and editing **Balakrishnan Viswanthan** Validation, Writing- review and editing

**DECLARATION OF INTERESTS**

The authors declare that they have no known competing financial interests or personal relationships that could have appeared to influence the work reported in this paper.

**REFERENCES**

[1] A. Nilsson and L.G.M. Pettersson, Nat. Commun. **6**, 1 (2015).
[2] P.G. Debenedetti and F.H. Stillinger, Nature **410**, 259 (2001).
[3] O. Mishima and H. E. Stanley, Nature **396**, 329 (1998).
[4] Z. Zhang and X.Y. Liu, Chem. Soc. Rev. **47**, 7116 (2018).
[5] P.G. Debenedetti, J. Phys. Condens. Matter **15**, R1669 (2003).
[6] C.A. Angell, Chem. Rev. **102**, 2627 (2002).
[7] J.M. Montes de Oca, S.R. Accordino, G.A. Appignanesi, P.H. Handle, and F. Sciortino, J. Chem. Phys. **150**, 144505 (2019).
[8] M.G. Mazza, N. Giovambattista, F.W. Starr and H.E. Stanley, Phys. Rev. Lett., 96, 057803 (2006)
[9] F. Sciortino, P. Gallo, P. Tartaglia, and S.H. Chen, Phys. Rev. E - Stat. Physics, Plasmas, Fluids, Relat. Interdiscip. Top. **54**, 6331 (1996).
[10] G. Adams and J.H.G. Gibbs, J. Chem. Phys. **43,** 139 (1965).
[11] A. Scala, F.W. Starr, E. La Nave, F. Sciortino and H.E. Stanley, Nature, **406**, 166 (2000).
[12] N. Giovambattista, S.V. Buldyrev, F.W. Starr and H.E. Stanley Phys. Rev. Lett. **90**, 085506 (2003).
[13] S. Karmakar, C. Dasgupta and S. Sastry, Rep. Prog. Phys. **79,** 016601 (2016).
[14] F. Klameth, P. Henritzi and M. Vogel, J. Chem. Phys. **140**, 144501 (2014); F Klameth and M. Vogel, J. Phys. Chem. Lett. **6**, 4385 (2015); J. Geske, M. Harrach, L. Heckmann, R. Horstmann, F. Klameth, N. Müller, E. Pafong, T. Wohlfromm, B. Drossel and M. Vogel, Z. Phys. Chem., **232**, 1187 (2018).
[15] J. C. Palmer, F. Martelli, Y. Liu, R. Car, A. Z. Panagiotopoulos and P. G. Debenedetti, Nature, **510**, 13505 (2014); G. Camisasca, D. Schlesinger, L. Zhovtobriukh, G. Pitsevich and L. G. M. Pettersson J. Chem. Phys. **151**, 034508 (2019).
[16] G. Camisasca, N. Galamba, K. T. Wikfeldt and L. G. M. Pettersson J. Chem. Phys. **150**, 224507 (2019).
[17] D. Chandler and J.P. Garrahan, Annu. Rev. Phys. Chem. **61**, 191 (2010).
[18] H. Tanaka, Faraday Discuss., **167**, 9 (2013).
[19] A. Widmer-Cooper, P. Harrowell, and H. Fynewever, Phys. Rev. Lett. **93**, 135701 (2004); A. Widmer-Cooper, P. Harrowell, and H. Fynewever, J. Chem. Phys. **126**,

154503 (2007).

[20] C. Balbuena, M.M. Gianeti and E.R. Soulé J. Chem. Phys. **149**, 094506 (2018)

[21] C. Balbuena and E.R. Soulé J. Phys.: Condens. Matter **32,** 045401 (2020)

[22] G.S. Matharoo, M.S.G. Razul, and P.H. Poole, Phys. Rev. E - Stat. Nonlinear, Soft Matter Phys. **74**, 050502 (2006).

[23] M.S.G. Razul, G.S. Matharoo, and P.H. Poole, J. Phys. Condens. Matter **23**, 235103 (2011).

[24] H.J.C. Berendsen, J.R. Grigera, and T.P. Straatsma, J. Phys. Chem. **91**, 6269 (1987).

[25] A.J. Easteal, W.E. Price, and L.A. Woolf, J. Chem. Soc. Faraday Trans. 1 Phys. Chem. Condens. Phases **85**, 1091 (1989).

[26] M.S.G. Razul and P.G. Kusalik, J. Chem. Phys. **134**, 014710 (2011).

[27] B. Hess, C. Kutzner, D. Van Der Spoel, and E. Lindahl, J. Chem. Theory Comput. **4**, 435 (2008).

[28] H.J.C. Berendsen, J.P.M. Postma, W.F. Van Gunsteren, A. Dinola, and J.R. Haak, J. Chem. Phys. **81**, 3684 (1984).

[29] M. Parrinello and A. Rahman, J. Appl. Phys. **52**, 7182 (1981).

[30] W.G. Hoover, Phys. Rev. A **31**, 1695 (1985).

[31] U. Essmann, L. Perera, M. L. Berkowitz, T. Darden, H. Lee, and L.G. Pedersen, J. Chem. Phys. **103**, 8577 (1995).

[32] P. Chau and A. J. Hardwick, Mol. Phys. **93**, 511 (1998).

[33] M. Vogel and S.C. Glotzer, Phys. Rev. Lett. 92, 255901 (2004)

[34] C. Donati, S.C. Glotzer, P.H. Poole, W. Kob, and S.J. Plimpton, Phys. Rev. E - Stat. Physics, Plasmas, Fluids, Relat. Interdiscip. Top. **60**, 3107 (1999).

[35] N. Giovambattista, S.V. Buldyrev, H.E. Stanley and F.W. Starr, Phys. Rev. E - Stat. Physics, Plasmas, Fluids, Relat. Interdiscip. Top. **72**, 011202 (2005).

[36] C. Marella and K. Muthukumarappan, in *Handb. Farm, Dairy Food Mach. Eng. Second Ed.* (Elsevier Inc., 2013), pp. 355–378.

[37] J. Toxopeus, V. Koštál, and B.J. Sinclair, Proc. R. Soc. B Biol. Sci. **286**, 20190050 (2019).

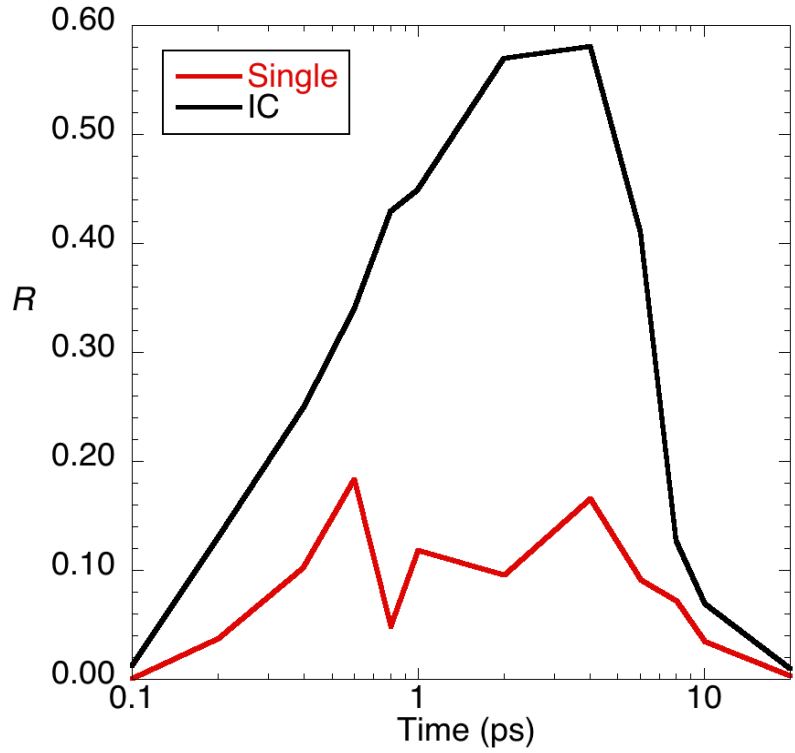

**Figure 1**. The system heterogeneities as shown by the correlation obtained by Pearson's correlation ($R$). The solid black line (IC) shows the correlation of $IM \langle r_i^2 \rangle_{ic} - O \langle sg_i \rangle_{ic}$ for an isoconfigurational run. The solid red line shows (Single) the $IM\, r_i^2 - O\, sg_i$ for the same starting configuration at equilibrium.

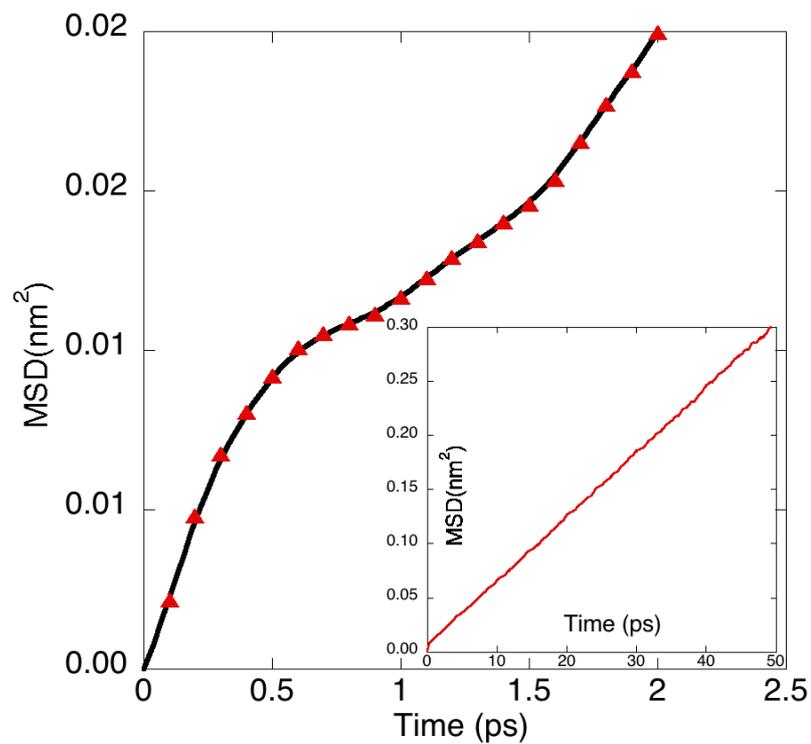

**Figure 2**. The mean-squared displacement (MSD) is shown as a function of time (up to 2.5ps) for $T = 265$ K. The weak caging effect for this temperature appears to start at 0.5 ps. The diffusive regime appears to start at 2 ps. The inset shows the MSD in the diffusive regime up to 50 ps.

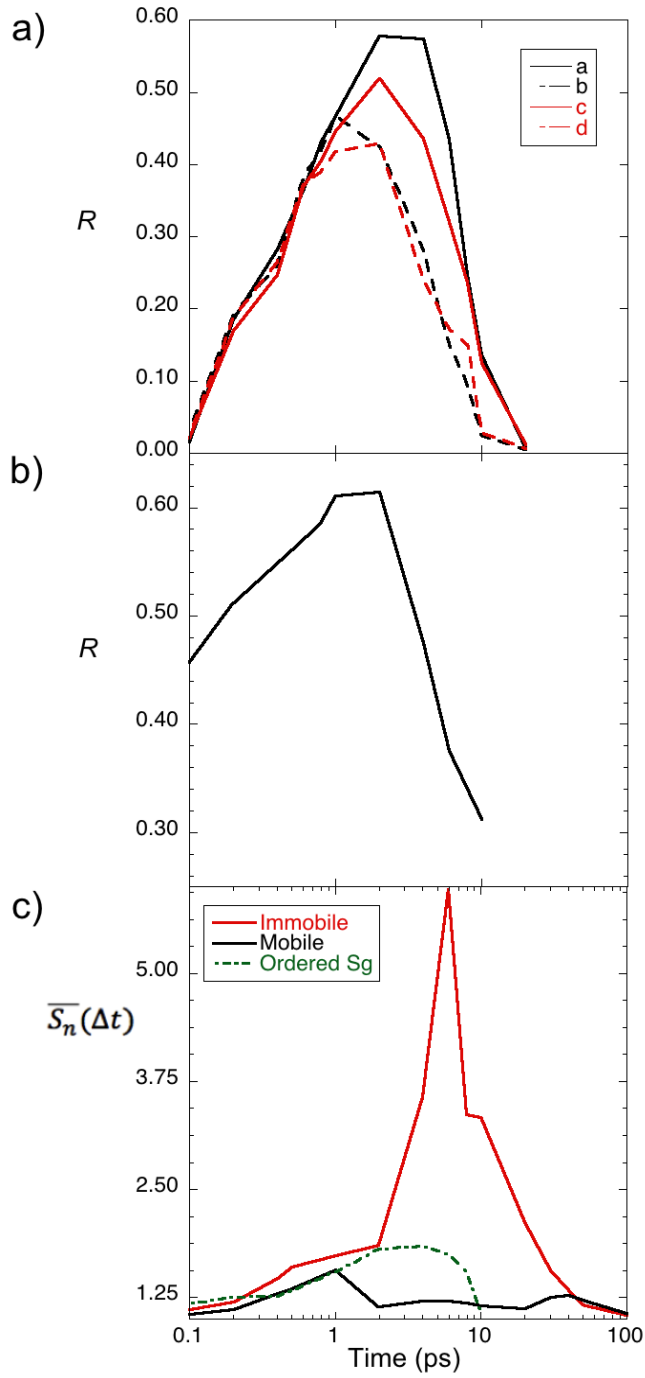

**Figure 3.** The heterogeneities that occur in the IC ensemble for various properties. **a)** The lines show the correlation obtained by Pearson's correlation, $R$: the solid black line shows the $IM \langle r_i^2 \rangle_{ic} - O \langle sg_i \rangle_{ic}$, the dotted black line shows $M \langle r_i^2 \rangle_{ic} - D \langle sg_i \rangle_{ic}$; the solid red line shows the $IM \langle r_i^2 \rangle_{ic} - O \langle sk_i \rangle_{ic}$ and the dotted red line shows $M \langle r_i^2 \rangle_{ic} - D \langle sk_i \rangle_{ic}$. **b)** The lines show the correlation obtained by Pearson's correlation, $R$ for $O \langle sg_i \rangle_{ic} - O \langle sk_i \rangle_{ic}$. **c)** Cluster data for the average weight cluster sizes, $\overline{S_n}(\Delta t)$ obtained during the time interval shown: the solid red line (Immobile) shows the immobile ($IM \langle r_i^2 \rangle_{ic}$) clusters; the black line (Mobile) shows the mobile ($M \langle r_i^2 \rangle_{ic}$) clusters; the dotted green line (Ordered Sg) shows the clusters of tetrahedrally ordered ($O \langle sg_i \rangle_{ic}$) clusters.

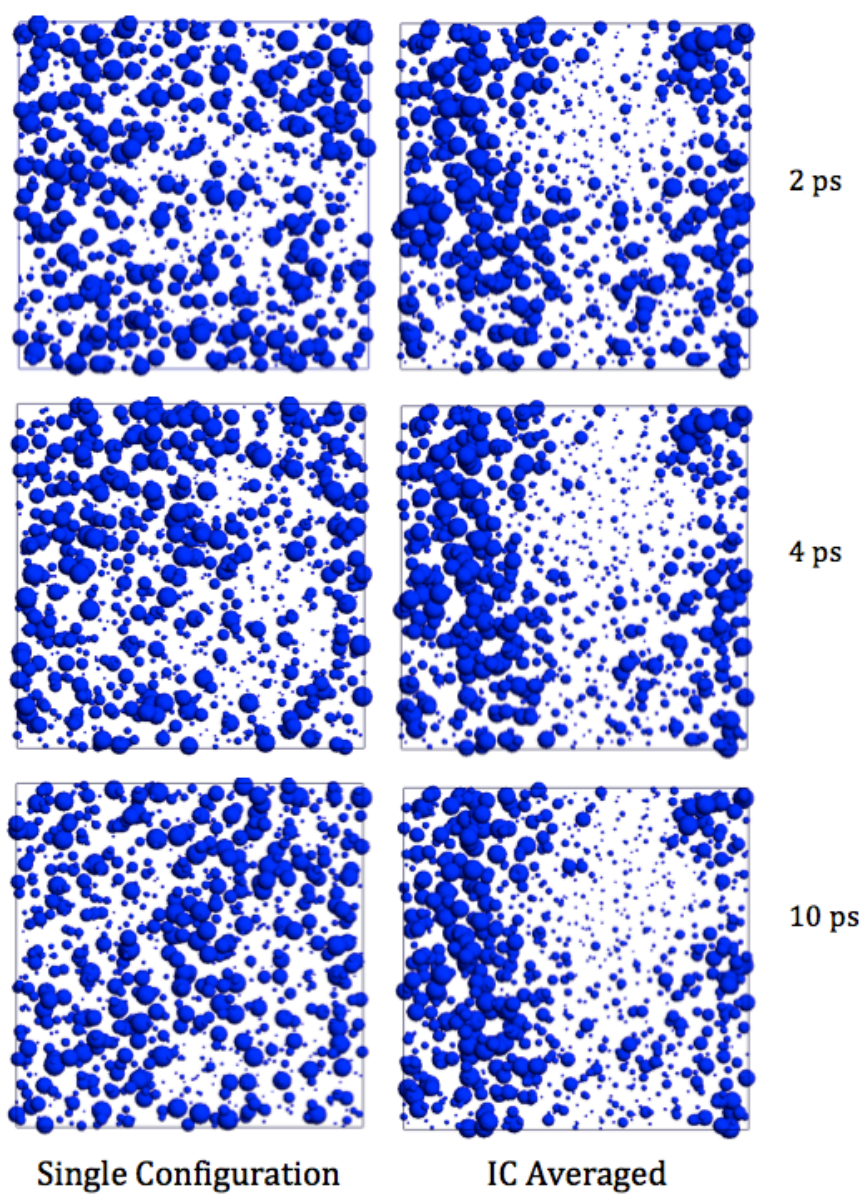

**Figure 4**. 3D-spatial plots representing the spatial variation and contrasting the $IM\ r_i^2$ (Single Configuration) and $IM\ \langle r_i^2 \rangle_{ic}$ (IC Averaged) from 2-10 ps. The plots are generated by first sorting for example $\langle r_i^2 \rangle_{ic}$ and assigning an integer rank $R_i$ from 1 to $N$, from smallest to largest. Each oxygen atomic position is then plotted as a sphere of radius $R_{min}\ exp\{[(R_i - N)/(1 - N)]\ log(R_{max}/R_{min})\}$, where $R_{max} = 0.1$ and $R_{min} = 0.01$. The result represents the rank of $\langle r_i^2 \rangle_{ic}$ or $r_i^2$ on an exponential scale, enabling spatial correlations to be visually represented, such that the largest blue spheres represent the most immobile O atoms. A particle is shown at its position in the initial configuration. The same starting configuration is shown. The single configuration displays little heterogeneity over time and appears to be random.

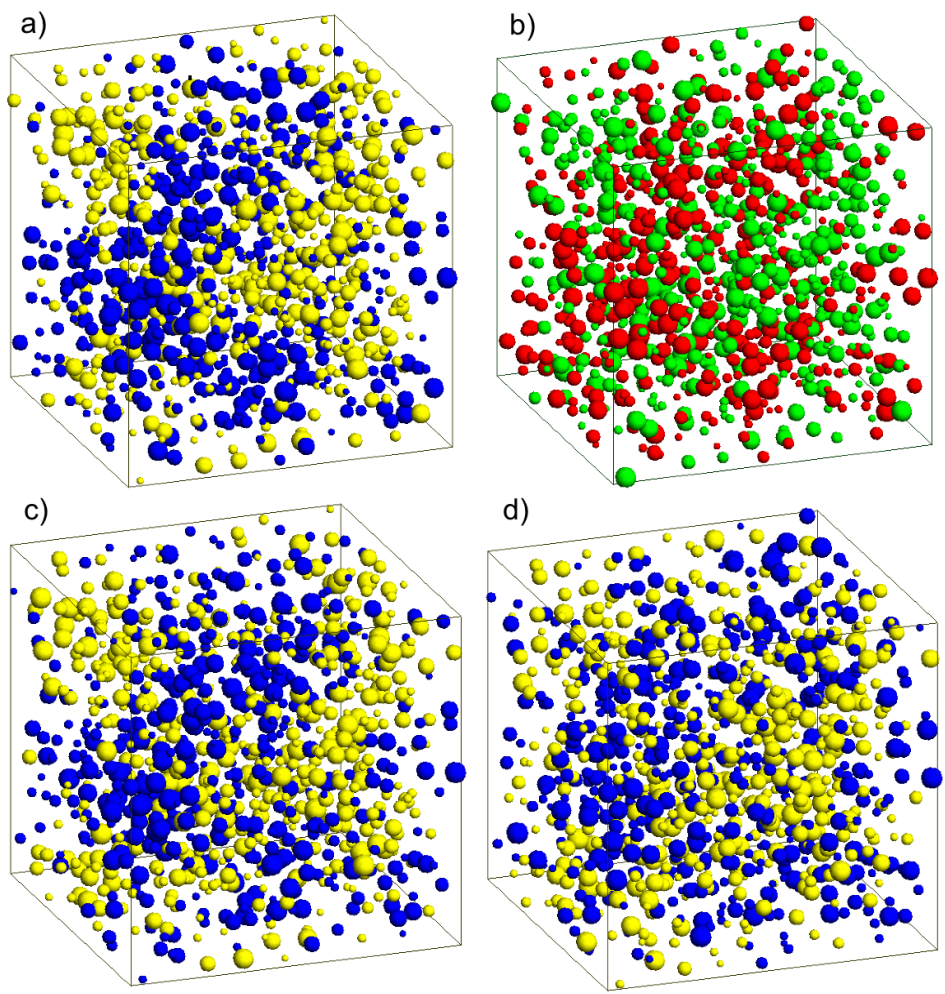

**Figure 5.** 3D-spatial plots representing the spatial variation of $IM \langle r_i^2 \rangle_{ic}$ (blue) and $M \langle r_i^2 \rangle_{ic}$ (yellow). These were obtained at **a)** 4 ps, **c)** 30 ps, **d)** 50 ps and **b)** $O \langle sg_i \rangle_{ic}$ (red) and $D \langle sg_i \rangle_{ic}$ (green) at 4 ps. This plot was generated the same way as in Figure 4. In this Figure the largest blue spheres represent the most immobile O atoms and the largest yellow spheres the most mobile, the largest red sphere the most tetrahedrally ordered and the largest green sphere the tetrahedrally disordered. A particle is shown at its position in the initial configuration of the IC ensemble.

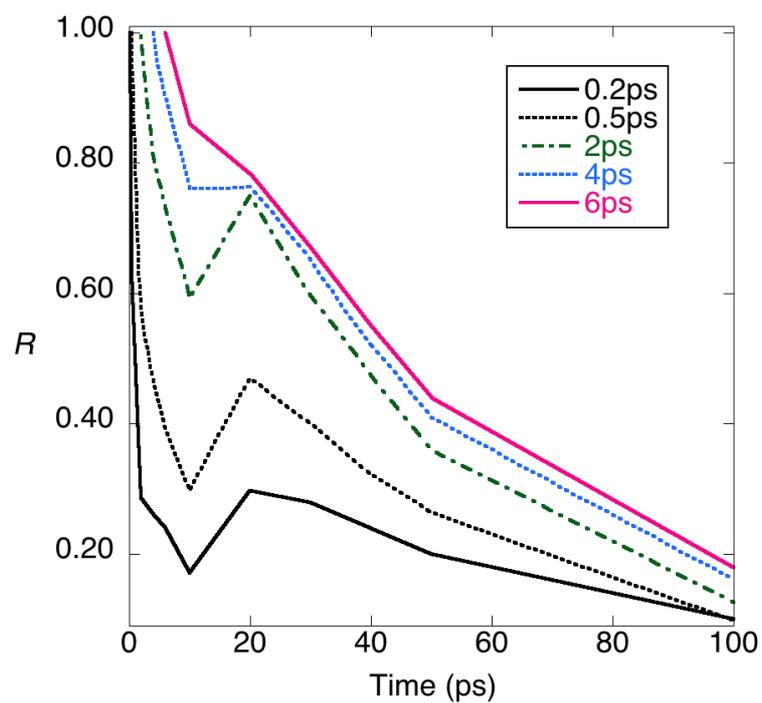

**Figure 6**. The Pearson's correlation, *R* data comparing dynamic propensities as a function of time. The propensities compared here are 0.2 ps (ballistic region of the MSD), 0.5 ps (caging region of the MSD) and 2, 4 and 6 ps (early diffusive region of the MSD).

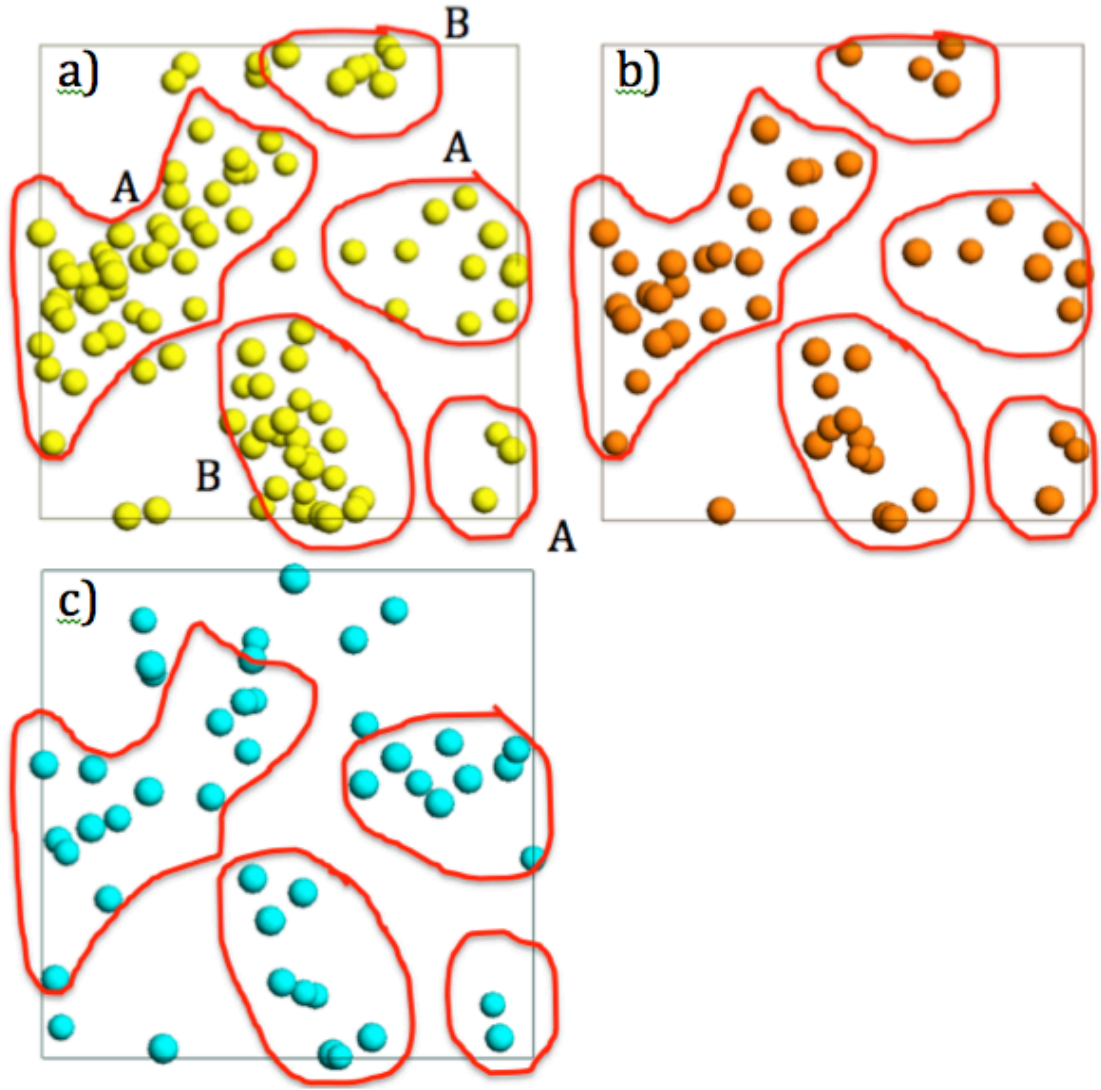

**Figure 7**. The 3D-spatial plots showing **a)** only the top 10% of $IM \langle r_i^2 \rangle_{ic}$ at 4 ps. The regions marked as *A* and *B* are extended regions in the simulation box (due to periodic boundary conditions). **b)** the top 10% of $IM \langle r_i^2 \rangle_{ic} - O \langle sg_i \rangle_{ic}$ at 4 ps in the same regions as a) **c)** the top 10% of $O \langle sg_i \rangle_{ic} - O \langle sk_i \rangle_{ic}$ at 1 ps in the same regions as a). The complete details for generating these plots are described in Fig. 3.